\documentclass[prb,twocolumn,superscriptaddress]{revtex4}
\usepackage{psfrag} 
\usepackage{graphicx}
\usepackage{dcolumn}
\usepackage{bm}
\usepackage{epsfig} 
\usepackage[isolatin]{inputenc} 
\usepackage{layout}
 
\usepackage{amssymb} 
\usepackage{amsmath}
 
\graphicspath{{images}{./bilder/}} 

\newcommand{\bra}[1]{\langle{#1}|}
\newcommand{\ket}[1]{|{#1}\rangle}
\newcommand{\braket}[2]{\langle{#1}|{#2}\rangle}

\newcommand{\ev}[1]{\langle{#1}\rangle}       
\newcommand{\eps}{\epsilon}
\newcommand{\al}{\alpha}
\newcommand{\euler}{\mbox{e}}  
\newcommand{\LichtPol}{\vec{\mathcal{E}}}  
\newcommand{\unu}{\boldsymbol{\nu}}   
\newcommand{\utnu}{\boldsymbol{\tilde{\nu}}}   
\newcommand{\tal}{\tilde{\alpha}}
\newcommand{\tbeta}{\tilde{\beta}}
\newcommand{\unull}{\mathbf{0}}   
\newcommand{\vD}{\vec{D}}
\newcommand{\cA}{\mathcal{A}}
\newcommand{\vd}{\vec{d}}
\newcommand{\cM}{\mathcal{M}}
\newcommand{\gi}{g_{\scriptscriptstyle I}}
\newcommand{\gr}{g_{\scriptscriptstyle R}}
\newcommand{\simpet}{\mbox{SP}}
\newcommand{\real}{\mbox{Re}}
\newcommand{\imag}{\mbox{Im}}

\newcommand{\vibre}{n_e}  
\newcommand{\vibrg}{n_g}  
\newcommand{\monham}{H_{\rm M}}  
\newcommand{\aggham}{H_{\rm A}}  
\newcommand{\vibIndex}{\lambda}
\newcommand{\Huang}{X}

\begin{document}
 
\title{The J- and H-bands of dye aggregate spectra: Analysis of the coherent exciton scattering (CES) approximation}

\author{Jan Roden}
\affiliation{%
Max-Planck-Institut f\"{u}r
Physik komplexer Systeme,
N\"othnitzer Str.\ 38,
D-01187 Dresden, Germany
}%
\affiliation{Theoretical Quantum Dynamics, Universit\"at Freiburg, 
Hermann-Herder-Str. 3, D-79104 Freiburg, Germany}
\author{Alexander Eisfeld}
\email{eisfeld@mpipks-dresden.mpg.de}
\affiliation{%
Max-Planck-Institut f\"{u}r
Physik komplexer Systeme,
N\"othnitzer Str.\ 38,
D-01187 Dresden, Germany
}%
\affiliation{Theoretical Quantum Dynamics, Universit\"at Freiburg, 
Hermann-Herder-Str. 3, D-79104 Freiburg, Germany}
\author{John S. Briggs}
\affiliation{Theoretical Quantum Dynamics, Universit\"at Freiburg, 
Hermann-Herder-Str. 3, D-79104 Freiburg, Germany}


%
%
%
%
%
\begin{abstract}

The validity of the CES approximation is investigated by comparison with direct diagonalisation of a model vibronic Hamiltonian of $N$ identical monomers interacting electronically.
Even for quite short aggregates ($N\gtrsim 6$) the CES approximation is shown to give
results in agreement with direct diagonalisation, for all coupling strengths,
except that of intermediate positive coupling (the H-band region). However,
previously excellent agreement of CES calculations and measured spectra
 in the H-band region was obtained [A. Eisfeld, J. S. Briggs, Chem. Phys. 324 (2006) 376] \cite{EiBr06_376_}. This  is shown to arise from use of the measured monomer spectrum which includes implicitly dissipative effects not present in the model calculation.

\end{abstract}
 

\maketitle

\section{Introduction}
In a series of papers
\cite{EiBr06_376_,EiBr02_61_,EiKnBr07_104904_,EiBr07_354_} the ``coherent
exciton scattering'' (CES) approximation was applied with some success to
calculate the absorption and circular dichroic spectra of a variety of 
aggregates of dye molecules. According to their structure, such aggregates
exhibit continuous J- or H-band absorption (see e.g.\ Refs.~\cite{Ko96__,AmVaGr00__}) and the CES approximation, working directly with the continuous vibronic absorption profile, is able to reproduce both types of bandshape in detail.
The CES approximation includes intra-molecular vibrations explicitly and coupling to other modes
only implicitly by working with the continuously-broadened experimental
spectra. Over the years there have been many
other approaches to the inclusion of vibronic effects on aggregate spectra, see for example Refs.~\cite{KnScFi84_481_,LuMu91_1588_,Ma98_195_,BaWaNe99_173_,BaWaNe00_403_,ScBrRe02_333_,SpZhMe04_10594_,SeMaEn06_354_,EgAl07_985_}.

The CES approximation is a mean-field type of approximation in which the exact monomer Green function, involving both electronic and vibrational degrees of freedom, is replaced by its average in the vibrational ground state. This allows the aggregate bandshape to be expressed in terms of the monomer absorption bandshape; the only fit parameter being the inter-monomer electronic interaction $V$. If $G(E)=(E-\monham-V+i\delta)^{-1}$ is the aggregate Green operator at energy $E$ and $g(E)=(E-\monham+i\delta)^{-1}$ that of the non-interacting monomers, one has the identity
\begin{equation}
        G=g+gVG
\end{equation}
where $\monham$ is the sum of monomer Hamiltonians and $V$ is the electronic interaction operator between monomers. The CES approximation corresponds to replacing the exact monomer Green operator $g$ by $\ev{g}$, where $\ev{\dots}$ denotes the aggregate vibrational ground state average.
Then
\begin{equation}
\label{exact_ev_G}
        \ev{G}=\ev{g}+\ev{gVG}
\end{equation}
becomes
\begin{equation}
\label{replace_g_by_ev_g}
        \ev{G}=\ev{g}+\ev{g}V\ev{G}
\end{equation}
when $V$ is assumed to be independent of vibrations. Eq.~(\ref{replace_g_by_ev_g}) has symbolically the solution
\begin{equation}
\label{symbolic_solution_for_G}
        \ev{G}=\frac{1}{1-\ev{g}V}\ev{g}.
\end{equation}
However, the physical content of the CES approximation is best seen by iterating Eq.~(\ref{replace_g_by_ev_g}) in a Born series, i.e.
\begin{equation}
\label{born_series_for_G}
        \ev{G}=\ev{g}+\ev{g}V\ev{g}+\ev{g}V\ev{g}V\ev{g}+\dots
\end{equation}
If the matrix element $G_{nm}=\bra{\pi_n}G\ket{\pi_m}$, where $\ket{\pi_n}$ is a state in which monomer $n$ is excited electronically with all other monomers in their ground electronic state, is taken, then Eq.~(\ref{born_series_for_G}) becomes
\begin{equation}
\label{born_series_for_G_in_pi_n_basis}
\begin{split}
\ev{G_{nm}}=&\ev{g_n}\delta_{nm}+\ev{g_n}V_{nm}\ev{g_m}\\
 &+\sum_{n'}\ev{g_n}V_{nn'}\ev{g_{n'}}V_{n'm}\ev{g_m}+\dots
\end{split}
\end{equation}
This shows that whenever electronic excitation is handed on from one monomer to another, the monomer which de-excites goes back into its \textit{ground vibrational state}. This is the essential ingredient in the CES approximation.

The CES approximation has been  used successfully  to describe the detailed lineshape of
various measured aggregate spectra
\cite{EiBr06_376_,EiBr02_61_,EiKnBr07_104904_,EiBr07_354_} in the intermediate
and strong coupling regime.
Following Simpson and Peterson \cite{SiPe57_588_} the strong and weak coupling
regimes are  defined by the conditions $V > \Delta$ or $V< \Delta$ respectively, where
$\Delta$ is the width of the monomer absorption spectrum. 
One speaks about intermediate coupling when  $V \approx \Delta$.

In some ways, the CES approximation performs better than it ought to, in that
in its original formulation \cite{BrHe70_1663_,BrHe71_865_} it involves only
the (zero temperature) aggregate ground state, represented as a simple product
of monomer vibrational ground states. 
However in the cases to which it has
been applied, measurements were made at room temperature in a variety of
solvents which also interact with the dye molecules. 
The origin of the success
of CES in this more general context may lie in the recent demonstration \cite{EiBr06_113003_} that its
essential structure (Eq.~(\ref{symbolic_solution_for_G})) is retained when the
average $\ev{\dots}$ is extended to include finite temperature and averages
over site solvent interactions. 
This question of temperature and solvent effects on J-aggregate spectra has
been studied extensively both experimentally \cite{ReWi97_7977_} and
theoretically \cite{HeMaKn05_177402_}. In the theoretical work  a purely
electronic Hamiltonian was used  for the aggregate but coupling to phonons of
the surroundings were considered to show that temperature effects can be
included successfully in this model. Here, since the CES method includes such
effects implicitly, coupling to phonons of the surrounding will not be
considered explicitly. Rather, we will include the intra-monomer vibrations
and compare the results of the CES approximation with those of a direct
diagonalisation of the aggregate vibronic Hamiltonian.

Although the original derivation \cite{BrHe70_1663_,BrHe71_865_} suggests that
the CES-approximation is applicable to strong coupling,
 in Ref.~\cite{LuFr77_36_} it was claimed that it  should only be valid for
 weak coupling.
Our subsequent comparison with experiment \cite{EiBr06_376_} indicates validity for both strong
and intermediate coupling.
To further throw light on these questions,
in this work the validity of the CES approximation will be examined in detail in its original form, in which an average is taken over the vibrational ground state only.
In section~\ref{sec:calc_of_aggr_spectra} the method of calculation is described. In section~\ref{sec:numerical_results} the results are presented and in section~\ref{sec:conclusion} our conclusions as to the limits of validity of the CES method are given.


\section{Theoretical method for calculation of aggregate spectra}
\label{sec:calc_of_aggr_spectra}

\subsection{Direct diagonalisation of the vibronic Hamiltonian}
\label{sec:exact_diag_of_vibr_ham}

We will consider an aggregate consisting of $N$ identical monomers having only one excited electronic state upon which vibrational manifolds are built.
For an arbitrary monomer $n$, the ground state is the product $\ket{\chi_n}\ket{\al_n}$ where $\ket{\chi_n}$ is the electronic ground state and $\ket{\al_n}$ denotes the vibrational state with quantum number $\al_n$ in this electronic state (we assume Born-Oppenheimer (BO) separability within a monomer).
In a corresponding notation the excited vibronic state of monomer $n$ is designated by the product $\ket{\phi_n}\ket{\beta_n}$.
Throughout this work $\chi$ and $\al$ will be used for the electronic {\it ground} state and its vibrational state and $\phi$ and $\beta$ for the {\it excited} electronic state and its vibrational state.

The aggregate Hamiltonian is the sum of the monomer Hamiltonians $H_n$ and the inter-monomer interaction operator $V$
\begin{equation}
\aggham=\monham+V=\sum_n H_n+V.
\end{equation} 
In the following it is assumed that the electron-electron interaction $V$ is
independent of nuclear co-ordinates. In previous work \cite{EiBr06_376_,EiBr02_61_,EiKnBr07_104904_,EiBr07_354_,BrHe70_1663_,BrHe71_865_} (and as used in Eq.~(\ref{born_series_for_G_in_pi_n_basis})) we defined one-exciton aggregate electronic states, where monomer $n$ is in the excited electronic state and all others in the ground state, as
\begin{equation}
\ket{\pi_n}=\ket{\chi_1}\ket{\chi_2}\cdots\ket{\phi_n}\cdots\ket{\chi_N}.
\end{equation}
Here we generalise this notation to define an aggregate vibronic state in which monomer $n$ is excited electronically and in vibrational state $\ket{\beta_n}$ and all other monomers in the ground electronic state with vibrational quantum number $\al_i$ for monomer $i$. This generalised basis is written
\begin{equation}
\label{vibronic_pi_basis}
\ket{\pi_n,\unu_n}=\ket{\pi_n}\ket{\al_1}\cdots\ket{\beta_n}\cdots\ket{\al_N}
\end{equation}
i.e. $\unu_n$ denotes the sequence of vibrational quantum numbers $(\al_1\dots\beta_n\dots\al_N)$. Note that this simple product basis is a BO basis for the aggregate as a whole. However, the inter-monomer interaction mixes these basis states to give no BO separation of the total aggregate eigenfunctions. In this chosen basis the matrix elements of the aggregate Hamiltonian are
\begin{equation}
\label{aggr_ham_pi_basis}
\begin{split}
\bra{\pi_n,\unu_n}&\aggham\ket{\pi_m,\utnu_m}\\
=&\eps^{\unu_n}
\delta_{nm}\delta_{\unu_n\utnu_m}
+V_{nm}\braket{\beta_n}{\tal_n}\braket{\al_m}{\tbeta_m}\prod^N_{\substack{i=1\\i\neq n,m}}\delta_{\al_i\tal_i}
\end{split}
\end{equation}
with 
\begin{equation}
\label{eq:eps^unu}
\eps^{\unu_n}\equiv\eps^{\al_1}+\dots+\eps^{\beta_n}+\dots+\eps^{\al_N}.
\end{equation}
Here $\eps^{\beta_n}$ is the energy of monomer $n$ in  the vibrational state
$\ket{\beta_n}$ of the excited  electronic state and $\eps^{\al_i}$ that of
monomer $i$ in the vibrational state $\ket{\al_i}$ of the electronic ground state. The quantum numbers $\tal_n$ and $\tbeta_m$ are elements of the sequence $\utnu_m$. The Franck-Condon (FC) factor $\braket{\al_n}{\beta_n}$ denotes the overlap of vibrational state $\beta_n$ of the excited electronic state with vibrational state $\al_n$ of the ground electronic state.

The eigenstates and eigenenergies of the aggregate are defined by
\begin{equation}
\aggham\ket{\Psi_l}=E_l\ket{\Psi_l}.
\end{equation}   
We designate by $\ket{\pi,\unull}$ the absolute vibronic aggregate ground state i.e.
\begin{equation}
\ket{\pi,\unull}=\prod^N_{i=1}\ket{\chi_i}\ket{\al_i=0}.
\end{equation}
Then the aggregate absorption cross-section energy dependence is given by the absorption strength
\begin{equation}
\label{aggr_absorp_cross-section}
\cA(E_l)=\left|\bra{\pi,\unull}\LichtPol\cdot\vD_{\rm A}\ket{\Psi_l}\right|^2
\end{equation}
where $\LichtPol$ is the light polarisation vector and the aggregate dipole
operator $\vD_{\rm A}$ is the sum of monomer operators $\vD_n$.
To evaluate Eq.~(\ref{aggr_absorp_cross-section}) further an identity is inserted  between $\vD_{\rm
  A}$ and $\ket{\Psi_l}$ so that 
\begin{equation}
\label{aggr_absorp_cross-section_with_identity}
\cA(E_l)=\left|\sum_{m,\unu_m}\LichtPol\cdot\bra{\pi,\unull}\sum_n\vD_{n}\ket{\pi_m,\unu_m}\braket{\pi_m,\unu_m}{\Psi_l}\right|^2
\end{equation}
The  dipole matrix element appearing in Eq.~(\ref{aggr_absorp_cross-section_with_identity}) can be
simplified to give
\begin{equation}
\begin{split}
\bra{\pi,\unull}\vD_{n}&\ket{\pi_m,\unu_m}\\
=&\bra{\al_n\!=\!0}\bra{\chi_n}\vD_n\ket{\phi_n}\ket{\beta_n}\
\delta_{nm}\delta_{\unu_m,(0\dots\beta_n\dots0)}.
\end{split}
\end{equation}
 Consistent with our assumption that $V$ is independent of vibrations, we make the Condon approximation that $\vD_n$ is independent of vibrations so that the dipole transition matrix element of monomer $n$ is
\begin{equation}
\bra{\al_n}\bra{\chi_n}\vD_n\ket{\phi_n}\ket{\beta_n}\equiv\vd_n\braket{\al_n}{\beta_n}.
\end{equation}
The absorption strength Eq.~(\ref{aggr_absorp_cross-section}) then reduces to
\begin{equation}
\label{abs_strength_pi_n_basis}
\cA(E_l)=\Big|\sum_{n,\beta_n}\LichtPol\cdot\vd_n\braket{\al_n\!=\!0}{\beta_n}\braket{\pi_n,(0\dots\beta_n\dots 0)}{\Psi_l}\Big|^2.
\end{equation}
The aggregate (stick) spectrum is then given by
\begin{equation}
\label{aggr_stick_spec}
\cA(E)=\sum_l\cA(E_l)\delta(E-E_l).
\end{equation}

The dimension of the basis Eq.~(\ref{vibronic_pi_basis}) and therefore the
number of terms in the summation Eq.~(\ref{aggr_stick_spec}) depends on the
number of vibrational states taken into account. If all monomers have $\vibrg$
vibrational states of the ground electronic  state and $\vibre$ vibrational
states of the excited electronic state, the dimension  is given by
\begin{equation}
\label{dim_vibronic_pi_basis}
\mbox{dim}=N\cdot\vibre\cdot\vibrg^{N-1}.
\end{equation}
Therefore the number of absorption sticks is already large (about 10000) for an aggregate of ten monomers with two vibrational states in each electronic state.
Note that the dimension depends crucially on the number $\vibrg$ of vibrational states of the electronic ground state.

In the special case that we assume that all $N$ monomers are identical and
arranged in a circle, the dimension of the matrix to be diagonalised can be reduced by a factor $N$ using a
transition to a delocalised excitonic basis rather than the localised basis
Eq.~(\ref{vibronic_pi_basis}). 
For simplicity we specialise $V$ to nearest-neighbour coupling.
How the delocalised states are constructed is illustrated first by considering the simple case of a cyclic trimer ($N=3$).
The excitonic basis consists of states $\ket{k,\unu_1}$, with $k=2\pi/3,4\pi/3,2\pi$, which are simply a superposition
\begin{equation}
\label{trimer_vibr_k_basis}
\begin{split}
\ket{k,\unu_1}=&\frac{1}{\sqrt{N}}\left[\euler^{ik}\ket{\pi_1,(\beta_1,\al_2,\al_3)}\right.\\
&+\euler^{2ik}\ket{\pi_2,(\al_3,\beta_1,\al_2)}\\
&\left.+\euler^{3ik}\ket{\pi_3,(\al_2,\al_3,\beta_1)}\right]
\end{split}
\end{equation} 
of the three possible shifts of the $\ket{\pi_1,\unu_1}$ state along the
cyclic aggregate.
\begin{figure}
\psfrag{monomer}{\hspace{-0.03\textwidth}\begin{minipage}[t]{2cm}\small monomer \\ \raisebox{0.8cm}{  number}\end{minipage} }
\psfrag{1}{\small $1$}
\psfrag{2}{\small $2$}
\psfrag{3}{\small $3$}
\psfrag{N}{\small $N$}
\psfrag{a_N}{\small $\alpha_N$}
\psfrag{b_1}{\small $\beta_1$}
\psfrag{a_2}{\small $\alpha_2$}
\psfrag{a_3}{\small $\alpha_3$}
\psfrag{a_N-1}{\small $\alpha_{N\!-\!1}$}
\psfrag{a_N-2}{\small $\alpha_{N\!-\!2}$}
\psfrag{S_2}{ $S_2$}
\includegraphics[width=0.95\columnwidth]{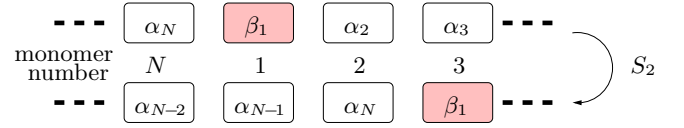}
\caption{Sketch of the action of the shift operator $S$. As example $S_2$ is considered. The excited electronic state is highlighted in gray. }
\label{shift_op}
\end{figure}
In general, for a cyclic $N$-mer the excitonic basis $\ket{k,\unu_1}$ is
\begin{equation}
\label{vibr_k_basis}
\ket{k,\unu_1}=\frac{1}{\sqrt{N}}\sum^N_{n=1}\euler^{ikn}S_{n-1}\ket{\pi_1,\unu_1},
\end{equation}
with $k=2\pi j/N$, $j=1,\dots ,N$ and where $S_m$ is a shift operator shifting both the electronic and vibrational excitation on the aggregate by $m$ monomers to a higher monomer index $i'=i+m$, i.e.
\begin{equation}
S_m\ket{\pi_1,\unu_1}=\ket{\pi_{1+m},(\al_{N+1-m}\dots\beta_1\dots\al_{N-m})}.
\end{equation}
The action of $S_m$ is also illustrated in Fig.~\ref{shift_op}.

Denoting by $W\equiv V_{n,n+1}$ the nearest-neighbour interaction between the monomers and defining
\begin{equation}
\label{def:C}
C\equiv
\begin{cases}
W & \mbox{for }N=2 \mbox{ (dimer)}\\
2W & \mbox{for }N>2
\end{cases}
\end{equation}
the matrix elements of $\aggham$ in the exciton basis Eq.~(\ref{vibr_k_basis}) are
\begin{equation}
\label{aggr_ham_vibr_k_basis}
\begin{split}
\bra{k,\unu_1}\aggham\ket{j,\utnu_1}=&\Big[\eps^{\unu_1}
\delta_{\unu_1\utnu_1}\\
&+\euler^{-ik}\frac{C}{2}\braket{\beta_1}{\tal_2}\braket{\al_N}{\tbeta_1}\prod^{N-1}_{i=2}\delta_{\al_i\tal_{i+1}}\\
&+\euler^{ik}\frac{C}{2}\braket{\beta_1}{\tal_N}\braket{\al_2}{\tbeta_1}\prod^{N-1}_{i=2}\delta_{\al_{i+1}\tal_i}\Big]\delta_{kj}.
\end{split}
\end{equation}
Note that 
$
\unu_1=(\beta_1,\al_2,\dots,\al_N)
$.
From Eq.~(\ref{aggr_ham_vibr_k_basis}) one sees that $\aggham$ is diagonal in the exciton $k$ basis i.e. the eigenvalue problem
\begin{equation}
\aggham\ket{\Psi_{k\vibIndex}}=E_{k\vibIndex}\ket{\Psi_{k\vibIndex}}
\end{equation}
can be solved separately for each $k= 2\pi/N,\dots ,2\pi$ with $\mbox{$\vibIndex=1,\dots,\vibre\cdot\vibrg^{N-1}$}$ ($\vibrg$ and $\vibre$ are the numbers of vibrational states taken into account in the ground and excited electronic state of each monomer, see Eq.~(\ref{dim_vibronic_pi_basis})). The absorption strength (see Eq.~(\ref{abs_strength_pi_n_basis})) from the aggregate ground state $\ket{\pi,\unull}$ to the eigenstate $\ket{\Psi_{k\vibIndex}}$ is given by
\begin{equation}
\begin{split}
\cA(E_{k\vibIndex})=&\frac{1}{N}\Big|\LichtPol\cdot\sum^N_{n=1}\euler^{ikn}\vd_n\Big|^2\\
&\times\left|\sum_{\beta_1=0}^{\vibre-1}\braket{\al_1=0}{\beta_1}\braket{k,(\beta_1,0\dots 0)}{\Psi_{k\vibIndex}}\right|^2.
\end{split}
\end{equation}
The first factor on the r.h.s. is the effective electronic dipole moment of the aggregate in exciton state $k$ and depends upon the aggregate geometry. In the simplest case where all monomer dipoles are parallel, this factor reduces to $N\left|\LichtPol\cdot\vd_1\right|^2\delta_{k,2\pi}$. In this case, to calculate the absorption spectrum, only the block $k=2\pi$ of the Hamiltonian matrix~(\ref{aggr_ham_vibr_k_basis}) needs to be diagonalised and the stick spectrum is given by (see Eq.~(\ref{aggr_stick_spec}))
\begin{equation}
\cA(E)=\sum_{\vibIndex}\cA(E_{2\pi,\vibIndex})\delta(E-E_{2\pi,\vibIndex})
\end{equation} 
with
\begin{equation}
\label{abs_num}
\cA(E_{2\pi,\vibIndex})=N\Big|\LichtPol\cdot\vd_1\!\!\!\sum_{\beta_1=0}^{\vibre-1}\braket{\al_1\!\!=\!\!0}{\beta_1}\braket{k,(\beta_1,0\dots 0)}{\Psi_{2\pi,\vibIndex}}\Big|^2.
\end{equation} 
Calculations of the aggregate spectrum from Eq.~(\ref{abs_num}) using the
eigenfunctions and eigenenergies from direct diagonalisation of the aggregate
Hamiltonian will be presented and referred to as DD calculations.

In the following discussion the absorption strength is given in arbitrary
units. All energies,including $\sigma$ and $\Delta$ are given in units of $\hbar\omega$.

%
\subsection{The monomer spectrum}

\begin{figure}[t]
\centering
\psfrag{a1}{}
\psfrag{d1}{\textbf{(a)}}
\psfrag{a2}{}
\psfrag{d2}{\textbf{(b)}}
\psfrag{a3}{}
\psfrag{d3}{\textbf{(c)}}
\psfrag{a4}{}
\psfrag{d4}{\textbf{(d)}}
\psfrag{absorption}{\hspace{-0.02\textwidth}Absorption [arb. u.]}
\psfrag{energy}{\hspace{-0.12\textwidth}\raisebox{-0.02\textwidth}{Energy $[\hbar\omega]$}}
\psfrag{delta0.1}{\small $\Delta=0.1$}
\psfrag{delta0.3}{\small $\Delta=0.3$}
\psfrag{delta0.8}{\small $\Delta=0.8$}
\psfrag{delta1.5}{\small $\Delta=1.5$}
\psfragscanon
\includegraphics[width=0.8\columnwidth]{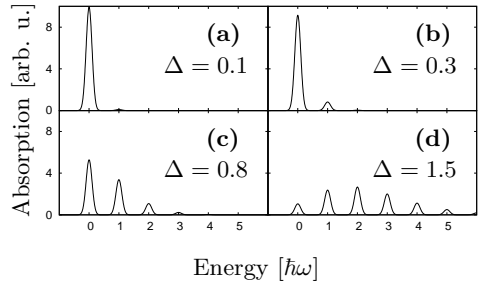}
\caption{Monomer spectra of different width (standard deviation) $\Delta$
  convoluted with narrow Gaussians (of standard deviation $\sigma=0.1$). All
  energies are in units of $\hbar\omega$.}
\label{fig:monomer_b0.1bis1.5u8o9sig0.1}
\end{figure}

All that is required to specify the vibrational shape of the monomer spectrum
 and carry through the calculation of the aggregate spectrum are the monomer
 Franck-Condon factors between vibrational states of upper and lower
 electronic states (for the definition see the discussion after Eq.~(\ref{eq:eps^unu})). Hence the calculations can be made for arbitrary ground
 and excited state BO potential surfaces. However, to conform with a large
 body of work, particularly using the second quantisation formalism, we will
 consider here that upper and lower potentials are both harmonic of the same
 curvature, with a shift $\overline{Q}$ of the minimum of the upper surface
 w.r.t.\ the minimum of the lower. Here $Q$ is the vibrational coordinate
 giving rise to a single dominant vibrational progression in the monomer
 spectrum. Despite its simplicity this model gives a surprisingly good fit to
 the spectra of a large variety of organic molecules. It has the advantage
 that the FC factors can be calculated analytically \cite{MeOs95__} and the monomer spectrum  obtained in a simple analytic form.
 Denoting the harmonic vibration frequency by $\omega$ and  defining the
 Huang-Rhys factor \cite{MeOs95__} by
\begin{equation}
\label{def:Huang}
X=\frac{\omega}{2\hbar}\overline{Q}^2,
\end{equation}  
the absorption strength of the monomer from the ground state (where $\al\!=\!0$), which is of relevance to the CES approximation, is
\begin{equation}
\label{monomer_abs_strength}
\cM(E)=\sum^{\vibre-1}_{\beta=0}\frac{X^{\beta}\euler^{-X}}{\beta !}\delta(E-\beta\hbar\omega).
\end{equation}
The width (standard deviation) of the monomer spectrum in this case of a
Poissonian distribution is given by
\begin{equation}
\Delta=\sqrt{X}\hbar\omega.
\end{equation}
Examples of monomer spectra for various values of the parameter $\Delta$ are
shown in Fig.~\ref{fig:monomer_b0.1bis1.5u8o9sig0.1}.
As is well-known, for small $\Huang$, i.e.\ small $\Delta$, the Poisson distribution gives only one dominant peak (Fig.~\ref{fig:monomer_b0.1bis1.5u8o9sig0.1}(a)) but becomes progressively more Gaussian as $\Huang$ increases and the progression broadens.

\subsection{The CES approximation}

It is clear from the discussion of the introduction that the CES approximation is equivalent to performing the diagonalisation of section~\ref{sec:exact_diag_of_vibr_ham} in a basis in which in the ground electronic state only the ground vibrational state is included. 
Then transfer of excitation to or from a monomer must proceed through this state as can be seen from the expansion~(\ref{born_series_for_G_in_pi_n_basis}) of the CES approximation. However the effect of the CES approximation on the absorption lineshape is better illustrated by the form~(\ref{symbolic_solution_for_G}).
The imaginary part of the monomer function $\ev{g(E)}$ is proportional to the monomer absorption strength Eq.~(\ref{monomer_abs_strength}).
 Once $\imag\ev{g}\equiv \gi$ is determined, the real part can be calculated from the Kramers-Kronig dispersion relation. Then the aggregate absorption strength can be calculated from $\imag\ev{G}$, where $\ev{G}$ is given by expression~(\ref{symbolic_solution_for_G}).

\begin{figure}[t]
\centering
\psfrag{E}{\Large $E$}
\psfrag{g}{\Large $\! \! \! \! \!  \! \! \!\displaystyle \frac{1}{\gr(E)}$}
\psfragscanon
\includegraphics[width=0.75\columnwidth]{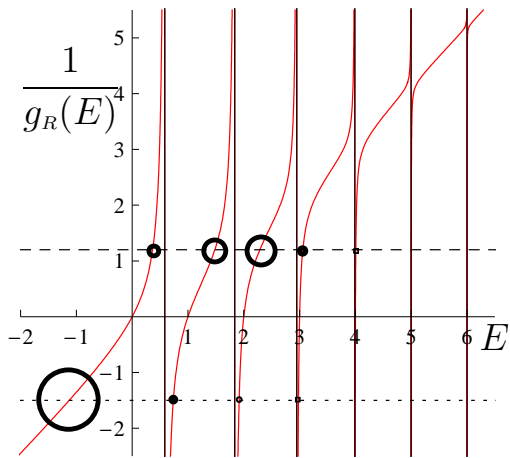}
\caption{ Monomer function $1/\gr$ as a function of energy $E$. The width of the monomer spectrum is
$\Delta=0.8$.
The absorption strength at each pole is indicated by the size of the circle.
The dashed horizontal  lines are for energies $C=-1.5$ and $C=1.2$.
}
\label{fig:disp_b0.8}
\end{figure}

For a stick monomer spectrum, the $\real\ev{g}\equiv \gr$ is readily
calculated and the function $1/\gr(E)$ is shown in Fig.~\ref{fig:disp_b0.8} (see also Ref.~\cite{EiBr06_376_}).
This function diverges at the discrete energy points where the monomer absorbs
(given by the energy values $\beta\hbar\omega$). Outside of these points the
function $\ev{g}$ is wholly real so that the aggregate function $\ev{G}$ has
poles (and hence there is absorption) where the coupling strength
$C=1/\gr$ (for the definition of $C$ see Eq.~(\ref{def:C})). 
These poles are given by the intersection of horizontal lines
(representing different coupling strength $C$) with the function
$1/\gr(E)$. From Fig.~\ref{fig:disp_b0.8} one sees that the location of
aggregate absorption depends sensitively on the sign and magnitude of $C$. 
As discussed in detail in Ref.~\cite{EiBr06_376_} this explains the qualitative dissimilarity of J and H band spectra. At the aggregate absorption energies (poles) $E_l$, $\gi=0$ so the expression~(\ref{symbolic_solution_for_G}) is not defined. Nevertheless \cite{EiBr06_376_} one can calculate the absorption function of the aggregate as
\begin{equation}
\label{absorption_from_greensfunction}
\cA(E)=\sum_l\left(\frac{\partial \gr^{-1}}{\partial E}\right)^{-1}_{E=E_l}\delta(E-E_l).
\end{equation}     
This shows that the absorption strength at a particular pole is inversely proportional to the slope of $\gr^{-1}$ at that point.
Hence only the poles where $\gr^{-1}$ has gradient in the order of unity absorb appreciably and these points are denoted by circles in Fig.~\ref{fig:disp_b0.8}. One also notes that in the CES approximation the number of poles is limited by the number $\vibre$ of vibrational states included on the upper potential irrespective of the number $N$ of monomers considered.

\section{Numerical Results}
\label{sec:numerical_results}

All calculations, both direct diagonalisation (DD) and CES calculations yield
stick spectra. The CES results in fact were obtained from the DD calculation
by restricting to the ground vibrational state  only (i.e.\ $\vibrg\!=\!1$),
but it was checked that the results agreed with those obtained from
Eq.~(\ref{absorption_from_greensfunction}). To facilitate comparison of CES
and DD results the calculated stick spectra have been convoluted with a narrow
Gaussian of standard deviation $\sigma\!=\!0.1\,\hbar\omega$.
As can be seen in Fig.~\ref{fig:monomer_b0.1bis1.5u8o9sig0.1} this width is
small enough to clearly resolve the individual peaks of the monomer absorption spectrum.

\begin{figure}
\psfrag{a1}{{\bf (a)}}
\psfrag{d1}{}
\psfrag{a2}{{\bf (b)}}
\psfrag{d2}{}
\psfrag{a3}{{\bf (c)}}
\psfrag{d3}{}
\psfrag{a4}{{\bf (d)}}
\psfrag{d4}{}
\psfrag{energy}{\hspace{-0.2\columnwidth}Energy[$\hbar \omega$]}
\psfrag{absorption}{\hspace{-0.04\columnwidth}Absorption [arb. u.]}
\includegraphics[width=1.\columnwidth]{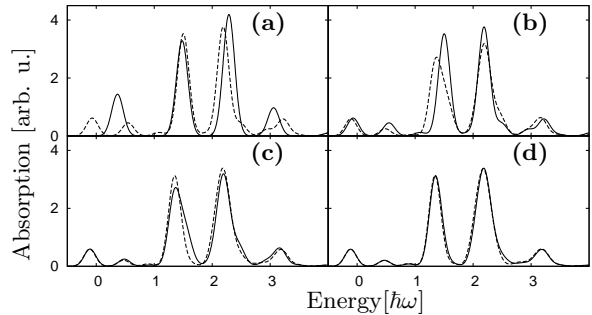}

\caption{\label{fig:convergenz}Illustration of the convergence criterion for
  the aggregate absorption. a) $n_g=1$ (solid) and $n_g=2$ (dashed). The overlap is
  68\%. b) $n_g=2$ (solid) and $n_g=3$ (dashed). The overlap is
  81\%.  c) $n_g=3$ (solid) and $n_g=4$ (dashed). The overlap is
  90\%. b) $n_g=4$ (solid) and $n_g=5$ (dashed). The overlap is
  95\%. Used parameters are $N=4$, $C=1.2$, $\Delta=0.8$, $n_e=9$. }
\end{figure}

The dimension of the Hamiltonian matrix (see Eq.~(\ref{aggr_ham_vibr_k_basis})) is $\vibre\cdot\vibrg^{N-1}$ and
computer time for diagonalisation therefore escalates, particularly as
$\vibrg$ increases. However all  DD spectra have been checked for convergence
and the cases presented have a minimum $95\,\%$ overlap with the spectra with
$\vibrg$ and/or $\vibre$ increased by unity. For all spectra shown, we have
used  $\vibre=9$.
In Fig.~\ref{fig:convergenz} our convergence criterion is illustrated for an
typical example.

There are three parameters deciding the shape of the aggregate spectrum,
namely, the coupling strength $C$, the monomer spectral width $\Delta$ and the
size $N$ of the aggregate. In the comparisons presented, representative points
in this 3-dimensional space are chosen. To some extent parameters $C$ and
$\Delta$ can be combined into the Simpson-Peterson ($\simpet$) parameter
$|C|/\Delta$, where strong coupling has $|C|/\Delta>1$ and weak coupling
$|C|/\Delta<1$. 
Although this is a very rough measure, it will emerge that,
for $N \gg 2$, the CES approximation is good for both weak and strong coupling
and only may have problems in the intermediate positive regime. 
Similarly, it has been
shown that the CES approximation is not good for dimers ($N=2$), since
electronic excitation is trapped indefinitely on just two monomers (see
Ref.~\cite{BrHe72_203_}). This conclusion is reinforced by the results
presented here, except in the case of weak coupling.

In considering these comparisons, two sum rules satisfied by both DD and CES spectra \cite{BrHe67_159_,BrHe70_1663_} should be kept in mind. One is that the total absorption strength of any spectrum is equal to that of the monomer (first sum rule SR1) and secondly that the mean energy of the spectrum (first moment) is shifted from that of the monomer by exactly the coupling energy $C$ (second sum rule SR2).

Since the CES result is independent of $N$, it is logical to present
 comparisons of DD spectra with CES for fixed values of $C$ and $\Delta$ and
 see at which $N$ the CES approximation becomes adequate.

Although we have performed calculations for monomer spectra with $\Huang$,
defined in Eq.~(\ref{def:Huang}),
ranging from $\Huang=0$ to $\Huang=2$, in this work we will focus on the
case $\Huang=0.64$, i.e.~$\Delta=0.8\hbar\omega$. This value of $\Huang$ is representative of those found for the dominant vibrational progression of many organic dyes.
\\

\textbf{1) Strong negative coupling}\\

\begin{figure}[t]
\centering
\psfrag{n2}{\hspace{-0.016\textwidth}\small $N=2$}
\psfrag{n4}{\hspace{-0.016\textwidth}\small $N=4$}
\psfrag{n6}{\hspace{-0.016\textwidth}\small $N=6$}
\psfrag{n8}{\hspace{-0.016\textwidth}\small $N=8$}
\psfrag{u8}{\hspace{-0.016\textwidth}\small $\vibrg=8$}
\psfrag{u4}{\hspace{-0.016\textwidth}\small $\vibrg=4$}
\psfrag{u2}{\hspace{-0.016\textwidth}\small $\vibrg=2$}
\psfrag{a1}{\hspace{-0.04\textwidth}\textbf{(a)}}
\psfrag{d1}{}
\psfrag{a2}{\hspace{-0.04\textwidth}\textbf{(b)}}
\psfrag{d2}{}
\psfrag{a3}{\hspace{-0.04\textwidth}\textbf{(c)}}
\psfrag{d3}{}
\psfrag{a4}{\hspace{-0.04\textwidth}\textbf{(d)}}
\psfrag{d4}{}
\psfrag{absorption}{\hspace{0.05\textwidth}Absorption [arb. u.]}
\psfrag{energy}{\hspace{-0.02\textwidth}Energy $[\hbar\omega]$}
\psfragscanon
\includegraphics[width=0.6\columnwidth]{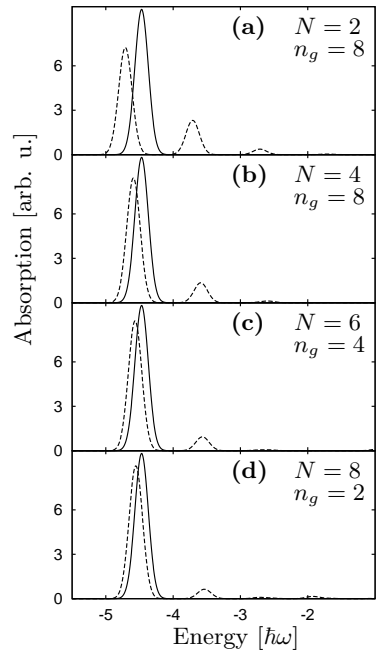}
\caption{ Strong negative coupling. Full curves CES spectra, dashed curves DD spectra, for $C=-5$, $\Delta=0.8$.
The values of $N$ and $\vibrg$ are indicated on the figures.}
\label{fig:spec_ces_n2bis8c-5.b0.8o9sig0.1}
\end{figure}

This case is the classic case of the formation of a J band. Representative
spectra are shown in Fig.~\ref{fig:spec_ces_n2bis8c-5.b0.8o9sig0.1} (for
$\simpet\approx 6$). The values of $\vibrg$ indicated on the figures are those
used in the calculation of the DD spectra. 
The results presented are for $C=\!-5$ but
we have checked that there is no change in the shape of the spectra for
$C\!<\!-5$. The monomer spectrum for this case is given in
Fig.~\ref{fig:monomer_b0.1bis1.5u8o9sig0.1}(c).  
As $N$ increases, essentially
only one strong peak on the low-energy side of the monomer band appears, whose mean energy is shifted by $C$ from the monomer mean energy. In Fig.~\ref{fig:spec_ces_n2bis8c-5.b0.8o9sig0.1}(d) ($N=8$) one sees that the CES approximation is in good agreement with the DD result.
As a result of the sum rules, the small peak near $E=-3.5$ in the DD spectrum
causes its J band to be slightly less intense and shifted to slightly larger
negative energy than the CES J band peak. 
On the dispersion curve of
Fig.~\ref{fig:disp_b0.8} one sees the origin of the J band as due to the
single pole at negative energy split off from the monomer band. This pole
carries all the oscillator strength. For the other poles the $\gr^{-1}$ curve is almost vertical, giving vanishing absorption according to Eq.~(\ref{absorption_from_greensfunction}).
\\

\textbf{2) Intermediate negative coupling}\\

\begin{figure}[t]
\centering
\psfrag{n2}{\hspace{-0.016\textwidth}\small $N=2$}
\psfrag{n4}{\hspace{-0.016\textwidth}\small $N=4$}
\psfrag{n6}{\hspace{-0.016\textwidth}\small $N=6$}
\psfrag{n8}{\hspace{-0.016\textwidth}\small $N=8$}
\psfrag{u8}{\hspace{-0.016\textwidth}\small $\vibrg=8$}
\psfrag{u4}{\hspace{-0.016\textwidth}\small $\vibrg=4$}
\psfrag{u2}{\hspace{-0.016\textwidth}\small $\vibrg=2$}
\psfrag{a1}{\hspace{-0.04\textwidth}\textbf{(a)}}
\psfrag{d1}{}
\psfrag{a2}{\hspace{-0.04\textwidth}\textbf{(b)}}
\psfrag{d2}{}
\psfrag{a3}{\hspace{-0.04\textwidth}\textbf{(c)}}
\psfrag{d3}{}
\psfrag{a4}{\hspace{-0.04\textwidth}\textbf{(d)}}
\psfrag{d4}{}
\psfrag{absorption}{\hspace{0.05\textwidth}Absorption [arb. u.]}
\psfrag{energy}{\hspace{-0.02\textwidth}Energy $[\hbar\omega]$}
\psfragscanon
\includegraphics[width=0.6\columnwidth]{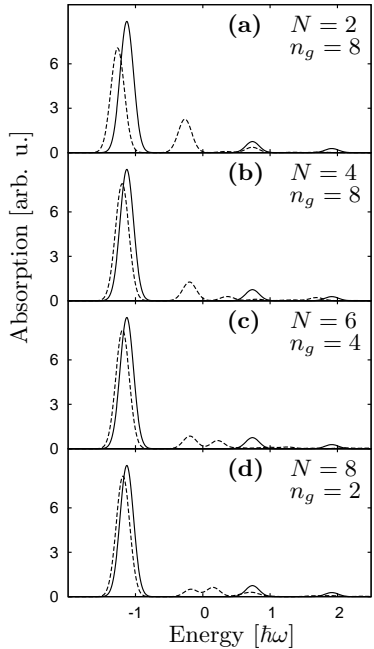}
\caption{ Intermediate negative coupling. Full curves CES spectra, dashed curves DD spectra, for $C=-1.5$, $\Delta=0.8$.
The values of $N$ and $\vibrg$ are indicated on the figures.}
\label{fig:spec_ces_n2bis8c-1.5b0.8o9sig0.1}
\end{figure}

Here we choose $C\!=\!-1.5$ so that $\simpet\approx 2$, all other parameters being as in Fig.~\ref{fig:spec_ces_n2bis8c-5.b0.8o9sig0.1}. From Fig.~\ref{fig:spec_ces_n2bis8c-1.5b0.8o9sig0.1} one notes that now the CES approximation has small subsidiary peaks.
As can be seen from Fig.~\ref{fig:disp_b0.8} these are due to additional poles in the monomer region acquiring some oscillator strength. Nevertheless the CES spectrum is dominated by the isolated-pole J-band peak. Although the peaks in the  region of monomer absorption are different in the DD calculation, these contributions diminish with increasing $N$ such that for $N=8$ there is again very good agreement between CES and DD results.
\\

\textbf{3) Weak negative and weak positive coupling}\\

\begin{figure}[tp]
\centering
\psfrag{n2}{}
\psfrag{u8}{}
\psfrag{a1}{}
\psfrag{d1}{}
\psfrag{absorption}{\hspace{-0.04\textwidth}Absorption [arb. u.]}
\psfrag{energy}{\hspace{-0.02\textwidth}Energy $[\hbar\omega]$}
\psfragscanon
\includegraphics[width=0.6\columnwidth]{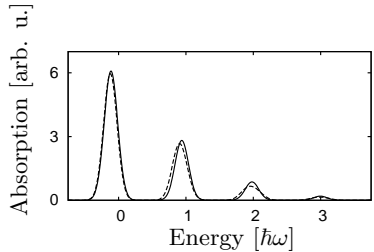}
\caption{ Weak negative coupling. Full curve CES spectrum, dashed curve DD spectrum, for $C=-0.2$, $\Delta=0.8$,
with $N=2$ and $\vibrg=8$.}
\label{fig:spec_ces_n2c-0.2b0.8o9sig0.1}
\end{figure}

\begin{figure}[tp]
\centering
\psfrag{n2}{}
\psfrag{u8}{}
\psfrag{a1}{}
\psfrag{d1}{}
\psfrag{absorption}{\hspace{-0.04\textwidth}Absorption [arb. u.]}
\psfrag{energy}{\hspace{-0.02\textwidth}Energy $[\hbar\omega]$}
\psfragscanon
\includegraphics[width=0.6\columnwidth]{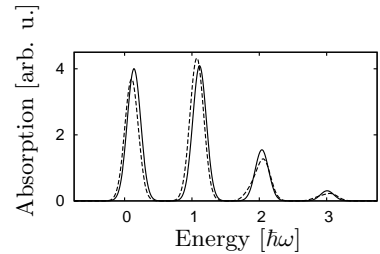}
\caption{ Weak positive coupling. Full curve CES spectrum, dashed
  curve DD spectrum, for $C=+0.3$ and $\Delta=0.8$ with $N=2$ and $\vibrg=8$.}
\label{fig:spec_ces_n2c0.3b0.8o9sig0.1}
\end{figure}

Here $C=-0.2$, $\simpet=0.25$ (Fig.~\ref{fig:spec_ces_n2c-0.2b0.8o9sig0.1}) and $C=+0.3$, $\simpet\approx 0.38$ (Fig.~\ref{fig:spec_ces_n2c0.3b0.8o9sig0.1}), all other parameters being taken as in Fig.~\ref{fig:spec_ces_n2bis8c-5.b0.8o9sig0.1}. 

In the case of weak coupling the DD spectrum does not change with $N$ for the case $\Delta=0.8$ and so only the dimer case $N=2$ is shown in Fig.~\ref{fig:spec_ces_n2c-0.2b0.8o9sig0.1} and Fig.~\ref{fig:spec_ces_n2c0.3b0.8o9sig0.1}.
Here there is excellent agreement, both in position and magnitude between DD and CES results. This is perhaps remarkable when one remembers that the DD results represent a convolution over up to about 10000 individual eigenvalues, whereas the CES spectrum arises from only the four main contributing poles for $C=-0.2$ and for $C=+0.3$, respectively.

To summarise, we have shown that, for aggregates of length $N\gtrsim 6$, the CES results, i.e.~including only the ground vibrational state in the electronic ground state, agree well with DD calculations where several vibrational states are included, for all negative coupling strength and for weak positive coupling.
For weak negative or positive coupling the CES method gives good results even for the dimer $N=2$.

The case of strong positive coupling is not shown, since if there were a strong shift to higher energies the absorption would overlap higher electronic bands and our model would break down.

There remains the most interesting case of intermediate positive coupling where the aggregate absorption occurs in the region of monomer absorption, typical of broad H band formation. This case will be considered in some detail.
\\

\textbf{4) Intermediate positive coupling}\\

\begin{figure}[tp]
\centering
\psfrag{n2}{\hspace{-0.016\textwidth}\small $N=2$}
\psfrag{n4}{\hspace{-0.016\textwidth}\small $N=4$}
\psfrag{n5}{\hspace{-0.016\textwidth}\small $N=5$}
\psfrag{n6}{\hspace{-0.016\textwidth}\small $N=6$}
\psfrag{u8}{\hspace{-0.016\textwidth}\small $\vibrg=8$}
\psfrag{u5}{\hspace{-0.016\textwidth}\small $\vibrg=5$}
\psfrag{u4}{\hspace{-0.016\textwidth}\small $\vibrg=4$}
\psfrag{a1}{\hspace{-0.04\textwidth}\textbf{(a)}}
\psfrag{d1}{}
\psfrag{a2}{\hspace{-0.04\textwidth}\textbf{(b)}}
\psfrag{d2}{}
\psfrag{a3}{\hspace{-0.04\textwidth}\textbf{(c)}}
\psfrag{d3}{}
\psfrag{a4}{\hspace{-0.04\textwidth}\textbf{(d)}}
\psfrag{d4}{}
\psfrag{absorption}{\hspace{0.05\textwidth}Absorption [arb. u.]}
\psfrag{energy}{\hspace{-0.02\textwidth}Energy $[\hbar\omega]$}
\psfragscanon
\includegraphics[width=0.6\columnwidth]{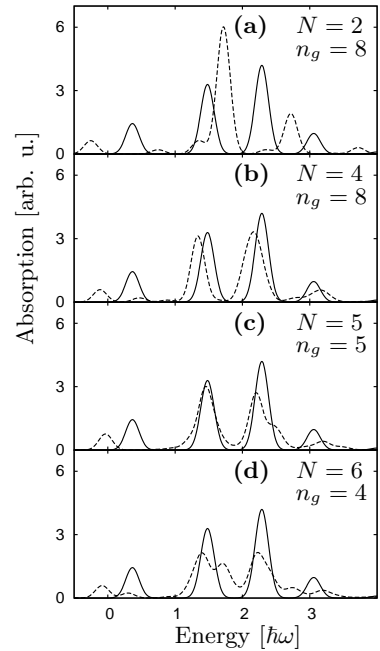}
\caption{Intermediate positive coupling. Full curves CES spectra, dashed curves DD spectra, for $C=+1.2$, $\Delta=0.8$, convoluted with a narrow Gaussian of standard deviation $\sigma=0.1$.}
\label{fig:spec_ces_n2bis6c1.2b0.8o9sig0.1}
\end{figure}

Note first that from Fig.~\ref{fig:disp_b0.8}, one sees that for intermediate
positive coupling, the aggregate poles in the CES approximation lie in the
region of monomer absorption and their strength and location then is very
sensitive to the coupling strength $C$.
In the following we consider the case $C\!=\!1.2$ (i.e.\ a $\simpet$ parameter
of $1.5$).
In Fig.~\ref{fig:disp_b0.8} the CES result for this case is illustrated (intersection of upper dashed horizontal line). 
In Fig.~\ref{fig:spec_ces_n2bis6c1.2b0.8o9sig0.1} a comparison of CES and DD spectra is shown.
\begin{figure}[tp]
\centering
\psfrag{n2}{\hspace{-0.016\textwidth}\small $N=2$}
\psfrag{n4}{\hspace{-0.016\textwidth}\small $N=4$}
\psfrag{n5}{\hspace{-0.016\textwidth}\small $N=5$}
\psfrag{n6}{\hspace{-0.016\textwidth}\small $N=6$}
\psfrag{u8}{\hspace{-0.016\textwidth}\small $\vibrg=8$}
\psfrag{u5}{\hspace{-0.016\textwidth}\small $\vibrg=5$}
\psfrag{u4}{\hspace{-0.016\textwidth}\small $\vibrg=4$}
\psfrag{a1}{\hspace{-0.04\textwidth}\textbf{(a)}}
\psfrag{d1}{}
\psfrag{a2}{\hspace{-0.04\textwidth}\textbf{(b)}}
\psfrag{d2}{}
\psfrag{a3}{\hspace{-0.04\textwidth}\textbf{(c)}}
\psfrag{d3}{}
\psfrag{a4}{\hspace{-0.04\textwidth}\textbf{(d)}}
\psfrag{d4}{}
\psfrag{absorption}{\hspace{0.05\textwidth}Absorption [arb. u.]}
\psfrag{energy}{\hspace{-0.02\textwidth}Energy $[\hbar\omega]$}
\psfragscanon
\includegraphics[width=0.6\columnwidth]{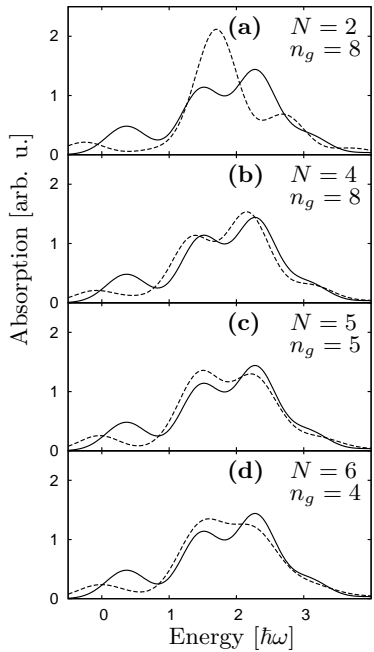}
\caption{Same as Fig.~\ref{fig:spec_ces_n2bis6c1.2b0.8o9sig0.1} but convoluted with a broad Gaussian with $\sigma=0.3$.}
\label{fig:spec_ces_n2bis6c1.2b0.8o9sig0.3}  
\end{figure}

One notes no good agreement of CES and DD results in this case, except to say that the dominant absorption is in the same spectral region for both methods.
Indeed in this case the DD results themselves vary significantly with the $N$ value and no convergence with $N$ is seen, in contrast to the case of intermediate negative coupling shown in Fig.~\ref{fig:spec_ces_n2bis8c-1.5b0.8o9sig0.1}.

\begin{figure}[t]
\centering
\psfrag{a1}{\textbf{(a)}}
\psfrag{d1}{}
\psfrag{a2}{\textbf{(b)}}
\psfrag{d2}{}
\psfrag{a3}{\textbf{(c)}}
\psfrag{d3}{}
\psfrag{a4}{\textbf{(d)}}
\psfrag{d4}{}
\psfrag{absorption}{\hspace{-0.12\textwidth}Absorption [arb. u.]}
\psfrag{energy}{\hspace{-0.02\textwidth}Energy $[\mbox{cm}^{-1}]$}
\psfragscanon
\includegraphics[width=0.6\columnwidth]{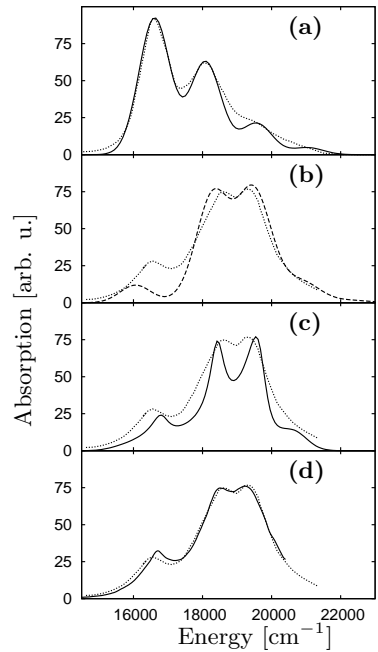}
\caption{ Dotted: measured Pinacyanol monomer (Fig.~(a))  and aggregate
  spectrum (Fig.~(b)-(d)) from Ref.~\cite{Sc38_1_} .
a) Solid line:  fit to a Poissonian convoluted with a Gaussian of width
  $\sigma\!=\!450\mbox{ cm}^{-1}$. 
b) Dashed: DD calculation with the peaks convoluted with the
Gaussian of Fig.(a). In the calculation $N=5$, $\vibrg=5$ was used.
c) Solid line:  CES spectrum obtained directly from the fitted monomer spectrum
of Fig.(a).
d) Solid line: CES spectrum obtained directly from the measured
monomer spectrum of Fig.(a).
}
\label{pina_scheibe_mon4.44_10-6_fit.eps}
\end{figure}

One must remember that the curves of Fig.~\ref{fig:spec_ces_n2bis6c1.2b0.8o9sig0.1} are calculated, in all cases, first as stick spectra and then convoluted with a Gaussian whose width is chosen arbitrarily as $\sigma=0.1$.
This width is considerably less than that typical of room--temperature solvent spectra.
Hence in Fig.~\ref{fig:spec_ces_n2bis6c1.2b0.8o9sig0.3}  the same data as in Fig.~\ref{fig:spec_ces_n2bis6c1.2b0.8o9sig0.1}  is shown convoluted with a Gaussian of $\sigma=0.3$ width.
Then one sees that, for $N\ge 4$, there is a much better, although still not exact agreement between CES and DD results.
Hence, even when averaged over a broad energy region, the CES method would appear not to perform well for the H-band case of intermediate positive coupling.
However, in Ref.~\cite{EiBr06_376_} we showed that the CES method gives excellent agreement, even in some fine details, with {\it measured} H-band spectra.
Important is that in Ref.~\cite{EiBr06_376_} the {\it measured} monomer spectrum, rather than a calculated Poissonian stick spectrum, was used as input to the CES method.
As we show now, it appears that this is crucial in order to obtain agreement with measured aggregate spectra and procedures based on fits to stick spectra in general appear less reliable.
To illustrate this, specifically we adopt the following strategy of comparing three different methods of calculation of the aggregate spectrum.

a) The measured monomer spectrum of Pinacyanol from Ref.~\cite{Sc38_1_}, which
was used in the CES calculations of Ref.~\cite{EiBr06_376_} and is shown in
Fig.~\ref{pina_scheibe_mon4.44_10-6_fit.eps}(a) (dotted line), is fitted to a stick
Poissonian distribution with each peak broadened by the same Gaussian. This
fitted spectrum is also shown in Fig.~\ref{pina_scheibe_mon4.44_10-6_fit.eps}(a) (solid line).
Taking $X$ and $\omega$ from the fitted monomer stick spectrum, the stick spectrum of the aggregate is calculated using the DD method. Then each peak is convoluted with the same Gaussian as used in the monomer spectrum.
The coupling strength $C$ has been optimised to obtain a best fit and the
overall calculated spectrum shifted by exactly the shift $S=480 \rm{ cm}^{-1}$
used in Ref.~\cite{EiBr06_376_} for the dye concentration considered.
The DD result for $N=5$ and $\vibrg=5$ is shown in Fig.~\ref{pina_scheibe_mon4.44_10-6_fit.eps}(b) and is in reasonable agreement with experiment, except for a large discrepancy in the peak around $16,500 \rm{ cm}^{-1}$.

b) The fitted continuous monomer spectrum of Fig.~\ref{pina_scheibe_mon4.44_10-6_fit.eps}(a) is used as input in
the CES calculation of the aggregate spectrum, again with $C$ optimised and
with $S=480 \mbox{ cm}^{-1}$.
Now, Fig.~\ref{pina_scheibe_mon4.44_10-6_fit.eps}(c), there is reasonable agreement in the peak position and
respective heights but the peaks appear too narrow.
This probably arises from the too-low absorption of the monomer fit in the
region around $19,000 \mbox{ cm}^{-1}$ (see  Fig.~\ref{pina_scheibe_mon4.44_10-6_fit.eps}(a)), leading to a too-low absorption in the
same region of the aggregate spectrum (as explained already, the local
aggregate absorption depends on the monomer absorption in the same locality).

c) From Ref.~\cite{EiBr06_376_}, the {\it measured} monomer spectrum is used as input
to the CES calculation of the aggregate spectrum.
The result is shown in Fig.~\ref{pina_scheibe_mon4.44_10-6_fit.eps}(d), where one notes excellent agreement with
the measured aggregate spectrum.

\section{Conclusion}
\label{sec:conclusion}

In its original form \cite{BrHe70_1663_,BrHe71_865_} the CES approximation is equivalent to the assumption that each monomer possesses only a ground vibrational state in the electronic ground state. 
Hence, in the aggregate only this state is involved when electronic excitation is transferred between monomers.
Here the validity of this approximation has been tested by comparing CES spectra with those obtained from direct diagonalisation (DD) of the aggregate vibronic Hamiltonian with a sufficient number of vibrational states in the electronic ground state to ensure convergence.
For negative values of the inter-monomer coupling energy, leading to a shift of the aggregate absorption to lower energies, good agreement of the CES spectra with DD spectra has been obtained for all negative coupling strengths, so long as the aggregate size $N$ exceeds $4$ or $5$ monomers. This success is probably due to the fact that for negative coupling, even intermediate values, a single dominant J-band peak splits off from the monomer region and carries most of the oscillator strength.
As shown in Fig.~\ref{fig:disp_b0.8} this mechanism of J-band formation is well-described by the CES approximation.

In the weak coupling regime, both negative and positive, the CES result agrees excellently with DD results since there is no strong mixing of monomer vibronic transitions.
In fact for weak coupling CES and DD results agree even for the dimer $N=2$.
The success of CES in weak coupling can be understood as follows.
For weak coupling one can truncate the Born series of the exact equation~(\ref{exact_ev_G}) to include the interaction $V$ only in lowest order, i.e.
\begin{equation}
\label{first_Born}
\ev{G}=\ev{g}+\ev{gVg}+\cdots
\end{equation}
Then one can show that, without any further approximation, the second term of the r.h.s. can be written
\begin{equation}
\ev{gVg}=\ev{g}V\ev{g}.
\end{equation}
Thus to lowest order in $V$ the exact equation~(\ref{first_Born}) is identical to the CES result Eq.~(\ref{born_series_for_G}).

Only for intermediate coupling, where the aggregate H-band absorption is shifted into the region of strong monomer absorption, do CES and DD results not agree, except qualitatively under low resolution.
This raises the question as to why the CES method applied previously \cite{EiBr06_376_} produces very good agreement with measured H-band spectra.
The answer has been shown to be that the procedure of calculating aggregate spectra using stick spectra with only one vibrational mode pro monomer taken into account and then fitting to a measured continuous spectrum by convoluting with Gaussians, is not a good strategy.
A good reproduction of aggregate H-band spectra is only obtained in the CES method when the {\it experimental} monomer spectrum is used.
Presumably this is because the DD model includes only one vibrational mode per monomer and  does not contain any influence of dissipation from the environment;
the eigenvalues are discrete.
When continuous experimental monomer spectra are used, such effects are included {\it implicitly} and the only assumption of the CES method in its generalised form \cite{EiBr06_113003_} is that such dissipative effects act in the same way on both monomer and aggregate spectra.

\vspace{0.7cm}

\begin{acknowledgements}
Financial support of this work and of that reported in Ref.~\cite{EiBr07_354_}
by the DFG in the project Br 728/11 is acknowledged gratefully. 
\end{acknowledgements}

\end{document}